\documentclass[12pt]{iopart}

\usepackage{graphicx}
\usepackage{bm}

\newcommand{\nwP}{{\tilde{P}}}
\newcommand{\bmx}{{\bm x}}
\newcommand{\bmv}{{\bm v}}
\newcommand{\bmk}{{\bm k}}
\newcommand{\bmq}{{\bm q}}
\newcommand{\et}{{\it et al}}

\newcommand{\simgt}{\lower.5ex\hbox{$\; \buildrel > \over \sim \;$}}
\newcommand{\simlt}{\lower.5ex\hbox{$\; \buildrel < \over \sim \;$}}

\begin{document}

\title[]
{Damping of the baryon acoustic oscillations in the 
matter power spectrum as a probe of the growth factor}

\author{Hidenori Nomura$^1$, Kazuhiro Yamamoto$^1$ and Takahiro Nishimichi$^2$}

\address{$^1$~Graduate School of Sciences, Hiroshima University, 
Higashi-Hiroshima, 735-8526,~Japan\\
$^2$~Department of Physics, School of Science, The University of Tokyo, 
Tokyo 113-0033, Japan}
\ead{hide@theo.phys.sci.hiroshima-u.ac.jp}

\begin{abstract}
We investigate the damping of the baryon acoustic oscillations (BAO) 
signature in the matter power spectrum due to the quasi-nonlinear 
clustering of density perturbations. 
On the basis of the third order perturbation theory, we construct a 
fitting formula of the damping in an analytic way. 
This demonstrates that the damping is closely related with the growth 
factor and the amplitude of the matter power spectrum. 
Then, we investigate the feasibility of constraining the growth factor 
through a measurement of the damping of the BAO signature.  
An extension of our formula including higher order corrections 
of density perturbations is also discussed. 
\end{abstract}

\maketitle

\section{Introduction}
The baryon acoustic oscillations (BAO) imprinted in the galaxy power
spectrum have recently attracted remarkable attention, as {a useful probe}
for exploring the origin of dark energy \cite{DETF,ESA}. 
The BAO signature has been clearly detected in the 2dFGRS and the SDSS
galaxy samples
\cite{Eisenstein,Huetsi,PercivalI,PercivalII,Tegmark}.
The feasibility of constraining the equation of state parameter of
dark energy $w$ is demonstrated \cite{PercivalIII,Okumura,HuetsiII}, 
where $w$ is defined by $w=p_{de}/\rho_{de}$, where $p_{de}$ and
$\rho_{de}$ are the pressure and the energy density of the dark energy
component, respectively. Furthermore, a lot of BAO survey projects are
under progress, or planned \cite{BOSS,FMOS,WFMOS,LSST,SKA,SPACE}. 

Though these BAO surveys will precisely measure the galaxy power spectra,
but we also need precise theoretical templates, in order
to obtain a useful cosmological constraint from observational data. 
In practice, we need to solve uncertainties about the gravitational 
nonlinear clustering of the density perturbations, the redshift-space
distortions, and the galaxy clustering bias, because these uncertainties
might yield some systematic effects in the BAO signature. 
Then, a lot of works related to these topics have been done recently, 
which are very important and challenging for future research of
dark energy \cite{Seo,SeoII,SeoIII,SeoIV,Angulo,AnguloII,Shoji,Takahashi,Wang,McDonald,Nishimichi}. 


In the present paper, we focus on the damping of the BAO signature in
the matter power spectrum due to the nonlinear gravitational
clustering. We investigate this damping in a semi-analytic way on the
basis of the third order perturbation theory of density fluctuations.
The nonlinear clustering is a consequence of mode-couplings
of the density fluctuations and the peculiar velocity divergence 
in Fourier space. 
Especially, we demonstrate how the damping of 
the BAO signature can be expressed with/without 
$P_{22}(k)$ and $P_{13}(k)$, which describe the mode-coupling 
(see sections 2 and 3). 
Then, we develop a simple fitting formula relevant to the damping,
which demonstrates the fact that the damping of the BAO signature 
is closely related with the growth factor and the amplitude of the 
matter power spectrum. This suggests that a measurement of the 
damping of the BAO signature might be a probe of the growth factor 
of the density fluctuations, though we need to further investigate 
if the effects of the redshift-space distortion or the galaxy 
clustering bias are influential to the damping.  

This paper is organized as follows; In section 2, we briefly review
the third order perturbation theory, and derive the second 
order power spectrum. In this formalism, the nonlinear gravitational 
clustering is described by the coupling of the Fourier modes of the 
density fluctuations and the peculiar velocity divergences.
In section 3, we investigate the damping of the BAO signature due to 
the coupling of the Fourier modes in an analytic way. 
And a fitting formula of the damping, applicable in weakly nonlinear 
regime, is developed. 
The formula is compared with a result of $N$-body simulation.
Extensions of the formula are also discussed. 
In section 4, with the use of the fitting formula, we demonstrate a
future feasibility of constraining the growth factor through a measurement of
the damping of the BAO signature. The last
section is devoted to a summary and conclusions. 
Throughout this paper, we use the unit in which the velocity of light
equals 1, and adopt the Hubble parameter $H_0=100h {\rm km/s/Mpc}$ 
with $h=0.7$, unless explicitly stated. 

\section{Quasi-nonlinear evolution of the matter power spectrum}
To investigate the effect of the nonlinear gravitational clustering in
the present work, we employ the standard perturbation theory (SPT) of
the density fluctuations and the peculiar velocity divergences 
\cite{Juszkiewicz,Vishniac,Fry,Goroff,Makino,Bertschinger,Bernardeau,Saito,Komatsu}.
Numerical simulation is a rigorous approach to the
gravitational clustering, but the perturbative approach has an advantage
to understand which factors are relevant to the damping of the BAO
signature in an analytic way. 
Recently, several authors have developed new formalism beyond the SPT
\cite{Valageas,ValageasII,Crocce,CrocceII,CrocceIII,McDonaldII,Matarrese,Izumi,Taruya,Matsubara}.
In the later subsection 3.3, we discuss a possible extension of our 
main result to
include these formalism. However, let us start with briefly reviewing the
derivation of the second order power spectrum on the basis of the SPT
up to the third order of perturbations.


We consider the matter fluctuations after the recombination whose
wavelength of interest is smaller than the horizon size, then the
evolution of the matter fluctuations can be analyzed by the 
pressure-less  nonrelativistic  fluid with the Newtonian gravity. 
Denoting the comoving coordinates by $\bmx$, and the conformal time by
$\eta$, the evolution equations are
\begin{eqnarray}
  \dot{\delta}(\bmx, \eta)+\nabla\cdot
    \left[\left(1+\delta(\bmx, \eta)\right)\bmv(\bmx, \eta)\right]=0,
\label{continuity}
\\
  \dot{\bmv}(\bmx, \eta)
    +\left(\bmv(\bmx, \eta)\cdot\nabla\right)\bmv(\bmx, \eta)
    +{\cal H}(\eta)\bmv(\bmx, \eta)=-\nabla\phi(\bmx, \eta),
\\
  \nabla^2\phi(\bmx, \eta)
    ={3\over 2}{\cal H}^2(\eta)\delta(\bmx, \eta),
\label{poisson}
\end{eqnarray}
where $\delta$ is the density contrast, $\bmv$ is the peculiar velocity,
$\phi$ is the gravitational potential, 
$a$ is the scale factor, a dot denotes the derivative with respect
to the conformal time, and ${\cal H}=\dot{a}/a$.
Here, the Einstein-de Sitter universe is assumed.

We ignore the rotational mode of the velocity, since our interest is 
only the growing solution, and the rotational mode is the decaying 
solution in the expanding universe. Then, we introduce
the velocity divergence, 
\begin{eqnarray}
  \theta(\bmx, \eta)\equiv \nabla\cdot\bmv(\bmx, \eta).
\end{eqnarray}
It is convenient to analyze the perturbations in the Fourier
space, and we define the Fourier coefficients as
\begin{eqnarray}
  \delta(\bmx, \eta)=\int {d^3k\over (2\pi)^{3}}
    \delta(\bmk, \eta)e^{-i\bmk\cdot\bmx},
\\
  \theta(\bmx, \eta)=\int {d^3k\over (2\pi)^{3}}
    \theta(\bmk, \eta)e^{-i\bmk\cdot\bmx}.
\end{eqnarray}
Then, equations \eref{continuity} - \eref{poisson} yield
\begin{eqnarray}
\fl  \hspace{10mm}\dot{\delta}(\bmk, \eta)+\theta(\bmk, \eta)=
    -\int d^3k_1 \int d^3k_2 \delta^{(3)}(\bmk_1+\bmk_2-\bmk)
    {\bmk\cdot\bmk_1\over k_1^2}
    \theta(\bmk_1, \eta)\delta(\bmk_2, \eta), 
\\
\fl  \hspace{10mm}\dot\theta(\bmk, \eta)+{\cal H}(\eta)\theta(\bmk, \eta)
    +{3\over 2}{\cal H}^2(\eta)\delta(\bmk, \eta)=
\nonumber \\
    -\int d^3k_1 \int d^3k_2 \delta^{(3)}(\bmk_1+\bmk_2-\bmk)
    {k^2\left(\bmk_1\cdot\bmk_2\right)\over 2k_1^2k_2^2}
    \theta(\bmk_1, \eta)\theta(\bmk_2, \eta),
\end{eqnarray}
where $\delta^{(3)}(\bmk)$ denotes the Dirac's delta function.
The right hand side of these equations describes the mode-couplings
which govern the nonlinear evolution of the matter fluctuations.

To solve these coupled equations, we adopt the perturbative expansion as
\begin{eqnarray}
  \delta(\bmk, \eta)=\sum_{n=1}^{\infty}a^n(\eta)\delta_n(\bmk), \qquad
    \theta(\bmk, \eta)
    ={\cal H}(\eta)\sum_{n=1}^{\infty}a^n(\eta)\theta_n(\bmk).
\end{eqnarray}
In general, the $n$-th order solution can be written as
\begin{eqnarray}
\fl \hspace{5mm} \delta_n(\bmk)=\int d^3q_1\cdots\int d^3q_n
    \delta^{(3)}\left(\sum_{i=1}^n\bmq_i-\bmk\right)
    F_n(\bmq_1,\ldots,\bmq_n)\prod_{i=1}^n\delta_1(\bmq_i),
\\
\fl \hspace{5mm} \theta_n(\bmk)=-\int d^3q_1\cdots\int d^3q_n
    \delta^{(3)}\left(\sum_{i=1}^n\bmq_i-\bmk\right)
    G_n(\bmq_1,\ldots,\bmq_n)\prod_{i=1}^n\delta_1(\bmq_i),
\end{eqnarray}
where $F_n(\bmq_1,\ldots,\bmq_n)$ and $G_n(\bmq_1,\ldots,\bmq_n)$ are
defined as
\begin{eqnarray}
\fl \hspace{12mm}F_n(\bmq_1,\ldots,\bmq_n)=
    \sum_{m=1}^{n-1}
    {G_m(\bmq_1,\ldots,\bmq_m)\over (2n+3)(n-1)}
    \left[(1+2n){\bmk\cdot\bmk_1\over k_1^2}
    F_{n-m}(\bmq_{m+1},\ldots,\bmq_n)\right.
\nonumber \\
    \hspace{47mm}\left.+{k^2(\bmk_1\cdot\bmk_2)\over k_1^2k_2^2}
    G_{n-m}(q_{m+1},\ldots,\bmq_n)\right],
\nonumber \\
\\
\fl \hspace{12mm}G_n(\bmq_1,\ldots,\bmq_n)=\sum_{m=1}^{n-1}
    {G_m(\bmq_1,\ldots,\bmq_m)\over (2n+3)(n-1)}
    \left[3{\bmk\cdot\bmk_1\over k_1^2}
    F_{n-m}(\bmq_{m+1},\ldots,\bmq_n)\right.
\nonumber \\
    \hspace{44mm}\left.+n{k^2(\bmk_1\cdot\bmk_2)\over k_1^2k_2^2}
    G_{n-m}(q_{m+1},\ldots,\bmq_n)\right].
\nonumber \\
\end{eqnarray}

Assuming that the first order density perturbations described by
$\delta_1(\bmk)$ is a Gaussian random field, we obtain the second order
matter power spectrum
\begin{eqnarray}
  P_{\rm SPT}(k, z)=D_1^2(z)P_{\rm lin}(k)+D_1^4(z)P_2(k),
\label{2nd}
\end{eqnarray}
where $P_{\rm lin}(k)$ is the linear power spectrum given by
\begin{eqnarray}
  (2\pi)^3\delta^{(3)}(\bmk+\bmk^{\prime})P_{\rm lin}(k)
    =\langle\delta_1(\bmk)\delta_1(\bmk^{\prime})\rangle,
\end{eqnarray}
$D_1(z)$ is the linear growth factor, and $P_2(k)$ is the second order
contribution to the power spectrum, which is conventionally expressed as
follows;
\begin{eqnarray}
  P_2(k)=P_{22}(k)+2P_{13}(k).
\label{p2}
\end{eqnarray}
Taking the 4-point correlations of $\delta_1(\bmk)$ into consideration,
we obtain the explicit form of $P_{22}(k)$ expressed as the
integral of the square of the linear power spectrum,
\begin{eqnarray}
  P_{22}(k)=2\int d^3q P_{\rm lin}(q)
    P_{\rm lin}\left(\left|\bmk-\bmq\right|\right)
    \left[F_2^{\rm s}(\bmq, \bmk-\bmq)\right]^2,
\label{defp22}
\end{eqnarray}
where $F_2^{\rm s}(\bmq_1, \bmq_2)$ is symmetrized over its arguments.
On the other hand, $P_{13}(k)$ has the form slightly different with
$P_{22}(k)$,
\begin{eqnarray}
  2P_{13}(k)=6P_{\rm lin}(k)\int d^3q P_{\rm lin}(q)
    F_3^{\rm s}(\bmq, -\bmq, \bmk).
\label{defp13}
\end{eqnarray}
The solid curve in figure 1 shows typical behaviour of $P_{22}(k)$ 
and $2P_{13}(k)$, where the cosmological parameters are 
$h=0.7$, ${\Omega}_m=0.28$, ${\Omega}_b=0.046$, $n_s=0.96$ and
$\sigma_8=0.82$. We discuss further details of the typical behaviour of
$P_{22}(k)$ and $2P_{13}(k)$ in the next section.

Originally, this perturbation expansion is consistently formulated 
in the Einstein de-Sitter universe. The result is extensively
used for the other general cosmological model, with the corresponding 
growth factor normalized as $D_1(z)=a(z)$ at $a(z)\ll1$. 
We follow this prescription because the validity of the extensive use 
of the result is known, for example, in the literature 
\cite{Komatsu,BernardeauII}.

\section{Damping of the BAO signature}
In this section, we will examine the damping of the BAO signature 
due to the nonlinear gravitational clustering with employing the
third order perturbation theory.
The BAO signature in the matter power spectrum can be extracted 
as follows:
\begin{eqnarray}
  B(k,z)\equiv{P(k,z)\over \tilde{P}(k,z)}-1,
\label{nlwiggle}
\end{eqnarray}
where $P(k,z)$ is the matter power spectrum including the BAO signature,
but $\tilde{P}(k,z)$ is the matter power spectrum without the BAO, which
is calculated using the no-wiggle transfer function in \cite{EH}. 
Hereafter, the quantity with the 'tilde' implies the quantity computed using
the no-wiggle transfer function. Within the third order perturbation
theory, we may write
\begin{eqnarray}
  \tilde P_{\rm SPT}(k, z)=D_1^2(z)\tilde P_{\rm lin}(k)+D_1^4(z)\tilde P_2(k),
\label{2ndtilde}
\end{eqnarray}
where $\tilde P_2(k)$ is 
\begin{eqnarray}
  \tilde P_2(k)=\tilde P_{22}(k)+2\tilde P_{13}(k),
\end{eqnarray}
and $\tilde P_{22}(k)$ and $\tilde P_{13}(k)$ are, respectively, 
defined by (\ref{defp22}) and (\ref{defp13}), but with the 
no-wiggle transfer function. 

Figure 2 shows $B_{\rm SPT}^{\rm exact}(k,z)$ as a function of the wavenumber 
$k$ for several redshifts, which is obtained using the third order 
perturbation theory (see also below). 
Here, the spatially flat universe with the cold dark matter (CDM) and 
the cosmological constant $\Lambda$ is assumed, where the cosmological
parameters are the same as those of Figure 1.
The oscillating behaviour of the curves is the BAO signature. As the redshift
becomes small, one can see that the amplitude of the oscillations 
decreases. This damping of the oscillations is more significant as 
$k$ is larger. 

\subsection{Analytic approach}
The aim of this paper is to understand the nature of the damping of
the BAO signature in detail.  
To this end, the analytic approach based on the third order perturbation
theory is useful. With the use of the formula of the second order
power spectra, (\ref{2nd}) and (\ref{2ndtilde}),  
we have
\begin{eqnarray}
  B_{\rm SPT}^{\rm exact}(k, z)
  ={P_{\rm SPT}(k,z)\over \tilde P_{\rm SPT}(k,z)}-1
  ={P_{\rm lin}(k)+D_1^2(z)P_2(k)\over 
    \tilde{P}_{\rm lin}(k)+D_1^2(z)\tilde{P}_2(k)}-1.
\label{defwnl}
\end{eqnarray}

First, we adopt  an approximation, 
\begin{eqnarray}
  P_2(k)\simeq \tilde{P}_{22}(k)+2P_{13}(k).
\label{appro22}
\end{eqnarray}
In brief,  $P_{22}(k)$ in (\ref{p2}) is replaced with $\tilde P_{22}(k)$.
The validity of this approximation is demonstrated in figure 1, where
the upper solid curve is $P_{22}(k)$, and the upper dotted curve is 
$\tilde P_{22}(k)$.
The validity of this approximation 
comes from the fact that the mode-coupling of different Fourier 
modes decrease the coherent BAO signature. 
As shown in this figure, the tiny oscillatory feature remains. 
This may induce somewhat the small shift of the peaks(troughs)
of the BAO, as mentioned in the reference \cite{CrocceIII}. 
However, this shift of peak location would not be problematic, as long 
as we are interested in the damping of the BAO signature. 

On the other hand, $P_{13}(k)$ can not be simply replaced with
$\tilde{P}_{13}(k)$. However, careful consideration leads to an
expression for $\tilde{P}_{13}(k)$ in terms of the linear power spectrum
multiplied by a monotonically decreasing function, as follows. 
First, we define 
\begin{eqnarray}
  B_{\rm lin}(k)\equiv {P_{\rm lin}(k)\over \tilde{P}_{\rm lin}(k)}-1,
\label{defwlin}
\end{eqnarray}
which corresponds to (\ref{defwnl}), within the linear theory of
density fluctuations. Note that $B_{\rm lin}(k)$ is not time-dependent. With
this definition, we obtain
\begin{eqnarray}
  2P_{13}(k)&=&6\tilde{P}_{\rm lin}(k)\left[1+B_{\rm lin}(k)\right]
    \int d^3q P_{\rm lin}(q) F_3^{\rm s}({\bf q}, -{\bf q}, {\bf k}).
\end{eqnarray}
Here, we apply the following approximation to the linear power spectrum
of the integrand, 
\begin{eqnarray}
  2P_{13}(k)&\simeq&6\tilde{P}_{\rm lin}(k)\left[1+B_{\rm lin}(k)\right]
    \int d^3q \tilde P_{\rm lin}(q) F_3^{\rm s}({\bf q}, -{\bf q}, {\bf k})
\nonumber
\\
  &=& 2\left[1+B_{\rm lin}(k)\right]\tilde{P}_{13}(k).
\label{appro13}
\end{eqnarray}
Substituting \eref{appro13} into \eref{appro22}, $P_2(k)$ is written as
\begin{eqnarray}
  P_{2}(k)=\tilde{P}_{2}(k)+2B_{\rm lin}(k)\tilde{P}_{13}(k).
\end{eqnarray}
Then, from \eref{defwnl}, we obtain
\begin{eqnarray}
  B_{\rm SPT}^{\rm exact}(k, z)\simeq{1
    +D_1^2(z)\displaystyle{2\tilde{P}_{13}(k)\over \tilde{P}_{\rm lin}(k)}\over
    1+D_1^2(z)\displaystyle{\tilde{P}_2(k)\over \tilde{P}_{\rm lin}(k)}}B_{\rm lin}(k).
\label{wnl1}
\end{eqnarray}
This formula indicates how the BAO signature is modified as the gravitational 
clustering evolves, which is expressed by the BAO signature in the linear
theory multiplied by the correction determined by the no-wiggle quantities
and the growth factor. 
The second term of the denominator in \eref{wnl1} is small 
within the perturbation scheme, we may expand it as
\begin{eqnarray}
  &&B_{\rm SPT}^{\rm exact}(k, z)\simeq
 \Biggl[1-D_1^2(z){\nwP_{22}(k)\over \nwP_{\rm lin}(k)}
    \Biggl\{1-D_1^2(z){\nwP_{2}(k)\over \nwP_{\rm lin}(k)}
\nonumber
\\
  &&\hspace{4cm}
  +D_1^4(z)\left({\nwP_{2}(k)\over \nwP_{\rm lin}(k)}\right)^2
    -\cdots\Biggr\}\Biggr]B_{\rm lin}(k),
\label{fullcorrection}
\end{eqnarray}
though the higher order terms make no sense because we are working in
the second order theory of the power spectrum. The expression 
\eref{fullcorrection} indicates that the leading effect of
the nonlinear mode-coupling on the damping is described by the factor 
$-D_1^2(z){\nwP_{22}(k)/\nwP_{\rm lin}(k)}$, and that 
the sign of the term clearly shows that this effect is a damping.

\subsection{Fitting formula}
As we are working within the third order perturbation theory, 
its prediction does not perfectly coincide with 
the result of full order computation, which can be obtained by 
$N$-body simulations. However, we believe 
that the prediction of the third order perturbation theory 
is useful in constructing a semi-analytic formula which 
reproduces the result of $N$-body simulations. 
Then, we here find a simple fitting formula which reproduces
the prediction of the third order perturbation theory. 
We discuss the validity of the formula in comparison with 
a $N$-body simulation in the below. 

Up to the second order of $D_1(z)$, \eref{fullcorrection} yields
\begin{eqnarray}
  B_{\rm SPT}^{\rm exact}(k, z)\simeq\left[1-D_1^2(z){\tilde{P}_{22}(k)\over
		\tilde{P}_{\rm lin}(k)}\right]B_{\rm lin}(k).
\label{leading}
\end{eqnarray}
{}From a detailed analysis of $\tilde P_{22}(k)/\tilde P_{\rm lin}(k)$ as a
function of $k$, we find that the following fitting formula works well,
\begin{eqnarray}
    {\tilde{P}_{22}(k)\over \tilde{P}_{\rm lin}(k)}
     =\sigma_8^2\left(k\over k_n\right)^2\left(1-{\gamma\over k}\right),
\label{fittingf}
\end{eqnarray}
where $\sigma_8$ is the rms matter density fluctuations averaged 
over the sphere with the radius of $8{h^{-1}}{\rm Mpc}$, $k_n$ 
and $\gamma$ are the constant parameters which depend on $\Omega_mh^2$, 
$\Omega_bh^2$ and $n_s$. Figures 3 and 4 show the best-fit value of 
$k_n$ and  $\gamma$ as a function of $\Omega_mh^2$ and $\Omega_bh^2$, 
where we fixed $n_s=0.96$. 
This shows that these two parameters depend 
on $\Omega_mh^2$ and $\Omega_bh^2$ linearly.
We can show a similar dependence on $n_s$. 
Then, we found the following fitting formula:
\begin{eqnarray}
  k_n=-1.03(\Omega_m h^2+0.077)(\Omega_b h^2-0.24)(n_s+0.92) ~~h{\rm Mpc}^{-1},
\\
  \gamma=-11.4(\Omega_m h^2-0.050)(\Omega_b h^2-0.076)(n_s-0.34) ~~h{\rm Mpc}^{-1}.
\end{eqnarray}
Though these dependence on the cosmological parameter 
might have to be investigated  more carefully, but the 
validity is guaranteed in the following narrow range
\begin{eqnarray}
  0.13\simlt\Omega_mh^2\simlt0.15,
\\
  0.022\simlt\Omega_bh^2\simlt0.024,
\\
  0.94\simlt n_s \simlt0.98.
\end{eqnarray}
In the present paper, we simply assume that the dependence on the 
other cosmological parameters except for 
$\Omega_m h^2$, $\Omega_b h^2$ and $n_s$ can be ignored. 

Finally, we obtain the heuristic expression of the leading correction 
for the BAO signature,
\begin{eqnarray}
  B_{\rm SPT}^{\rm fit}(k, z)=\left[1-f(\Omega_m h^2, \Omega_b h^2, n_s, \sigma_8, k, z)\right]
					B_{\rm lin}(k),
\label{finalfit}
\end{eqnarray}
with
\begin{eqnarray}
  f(\Omega_m h^2, \Omega_b h^2, n_s, \sigma_8, k, z)=
   \sigma_8^2 D_1^2(z) 
		\left(k\over k_n\right)^2\left(1-{\gamma\over k}\right).
\label{fit22}
\end{eqnarray}
Figure 5 demonstrates the agreement between this fitting formula, 
$B_{\rm SPT}^{\rm fit}$, and the prediction of the third order 
perturbation theory, $B_{\rm SPT}^{\rm exact}$, \eref{defwnl}. 
As one can see from this figure, the agreement becomes worse
as the redshift is lower and the wavenumber is larger.
Figure 6 shows the relative error of the fitting formula, 
$|B_{\rm SPT}^{\rm fit}-B_{\rm SPT}^{\rm exact}|/|B_{\rm SPT}^{\rm exact}|$, 
as a function of the redshift 
at the wavenumbers of P1,~P2,~P3,~T1,~T2 and T3, 
which are defined in Figure 5. 
The relative error is about less than $10~\%$ at worst in the 
range of the wavenumber $k\simlt0.2~{h}{\rm Mpc}^{-1}$ until
$z\sim 1$.
The sign of $B_{\rm SPT}^{\rm fit}-B_{\rm SPT}^{\rm exact}$ of 
the third peak (trough) changes around the
redshift 2.7 (1.4), where the relative error becomes zero.
For the redshift less than $1$, the agreement becomes
worse especially at the large wavenumber $k~(\simgt0.2~{h}{\rm Mpc}^{-1})$. 
However, we should
also note that the third order perturbation theory becomes worse
to reproduce $N$-body simulations for lower redshift and for larger
wavenumbers. 

Figure 7 shows a comparison of our fitting formula ($B_{SPT}^{\rm fit}$, 
solid curve) and a result of $N$-body simulation (squares with error bar) 
\cite{Nishimichi08}. This demonstrates that our fitting 
formula reproduces the result of $N$-body simulation for
$k\simlt0.2~{h}{\rm Mpc}^{-1}$ until $z\sim1$ within error bars, roughly.
In this conclusion, note that we focus on the damping of the BAO 
signature, not on the amplitude of the power spectrum itself.

In the $N$-body simulation, we adopt a $\Lambda$CDM
cosmology ($\Omega_m=0.279$, $\Omega_\Lambda=0.721$,
$\Omega_b/\Omega_m=0.165$, $h=0.701$, $n_s=0.96$, $\sigma_8=0.817$;
WMAP5 best fit value, \cite{WMAP5}), and
calculate the linear matter power spectrum using {\tt CAMB} \cite{CAMB}.
We adopt $512^3$ particles in periodic cubes with each side
$1000h^{-1}{\rm Mpc}$, and displace $N$-body particles using the second-order
Lagrangian perturbation theory (e.g., \cite{CrocceIV}) from uniform grid
positions at $z=31$. The simulations are carried out using the {\tt
Gadget2} code \cite{Springel} to output data of $4$ redshifts 
($z=3$, $2$, $1$, and $0.5$). 
Our total simulation volume is $8h^{-3}$Gpc$^{3}$, which  might be
small to investigate our scales of interest (first a few BAO peaks). 
Then we correct the deviations from the ideal case of
infinite volume, as follows. 

In addition to the time integration using {\tt Gadget2}, we also
  calculate the time evolution of the density contrast using the
  second-order perturbation theory starting from the identical initial
  condition with that used for the $N$-body simulation. We then calculate
  the power spectrum from the result of the perturbation theory, and
  measure the {\it deviation} from the the linear power spectrum at each
  output time. The deviation
  of our finite volume simulation is well explained by the 
  mode-couplings predicted by the second-order perturbation theory. 
  Then, we can obtain a corrected power spectrum by multiplying
  the measured power spectrum from $N$-body simulations by
  the ratio of the linear power spectrum to the power 
  spectrum predicted by the perturbation theory. Thus the
  error bars in the figure stand for the remaining standard errors
  after this correction, which are very small compared
  with the usual cases in the region of the small wavenumbers.  
  This is because the perturbation theory used for the correction 
  is more accurate at the small wavenumbers, 
  which can remove the deviations from the linear power spectrum better. 
  This correction does not improve the errors at large wavenumbers  
 (see \cite{Takahashi}, \cite{Nishimichi08} for more details).  
  Moreover, in order to
  construct the no-wiggle spectrum required for computing the ratio
  $B$, we use a cubic basis-spline fitting with the same break points
  as in \cite{Nishimichi} (see also \cite{PercivalIII}).


\subsection{Possible extension and discussion}
There are other possible effects which may affect the damping of the 
BAO signature:
the higher order nonlinear effect, the redshift-space distortions  and
the clustering bias. These effects may be influential to the damping of 
the BAO signature, however, there remains a lot of uncertainties about 
these effects, at present.
Here, let us consider a possible extension of our work to include
the higher order nonlinear corrections in real space.

The standard perturbation theory is useful to analyse the damping 
of the BAO signature in an analytic way, however, it is not enough 
for precise predictions that match with result of $N$-body 
simulations in the regime where the nonlinear effect becomes 
significant. Recently, several authors have developed 
non-perturbative approach to the nonlinear density clustering,
as mentioned above \cite{Valageas,ValageasII,Crocce,CrocceII,
CrocceIII,McDonaldII,Matarrese,Izumi,Taruya,Matsubara}.

As an alternative to the SPT, we here consider the work 
proposed by \cite{Matsubara}, which uses the technique 
of resumming infinite series of higher order perturbations 
on the basis of the Lagrangian perturbation theory (LPT). 
One of the advantage of this approach is the simplicity 
of the resulting expression of the nonlinear power 
spectrum, which enable us to incorporate the result
into our formula. 
In the framework of the LPT\cite{Matsubara}, 
the matter power spectrum can be given by
\begin{eqnarray}
&&P_{\rm LPT}(k,z)
=\rme^{-D_1(z)^2g(k)}\big[D_1(z)^2P_{\rm lin}(k)
\nonumber
\\
&&~~~~~~~~~~~~~~~~~~~~~~~~~~~~~~~~~~+D_1(z)^4P_2(k)
 +D_1(z)^4P_{\rm lin}(k)g(k)\big],
\label{lptpk}
\end{eqnarray}
where
\begin{eqnarray}
  g(k)={k^2\over 6\pi^2}\int dq P_{\rm lin}(q).
\end{eqnarray}
Corresponding to \eref{defwnl}, we define 
the BAO signature of the matter power spectrum based on the LPT,
\begin{eqnarray}
    B_{\rm LPT}^{\rm exact}(k, z)=
   {P_{\rm LPT}(k,z)\over \tilde P_{\rm LPT}(k,z)}-1,
\end{eqnarray}
where $\tilde P_{\rm LPT}(k,z)$ is defined by (\ref{lptpk}) but with 
the no-wiggle transfer function. 
Adopting an approximation, $g(k)\simeq \tilde{g}(k)$, we can 
obtain the following expression 
\begin{eqnarray}
    B_{\rm LPT}^{\rm exact}(k, z)\simeq{1+\displaystyle{
      {D_1^2(z)\over 1+D_1^2(z)\tilde{g}(k)}
      {2\tilde{P}_{13}(k)\over\tilde{P}_{\rm lin}(k)}}\over
    1+\displaystyle{
      {D_1^2(z)\over 1+D_1^2(z)\tilde{g}(k)}
      {\tilde{P}_2(k)\over \tilde{P}_{\rm lin}(k)}}}B_{\rm lin}(k).
\label{matsubara}
\end{eqnarray}
Repeating the same procedure from \eref{wnl1} to \eref{leading},
one can obtain the leading correction to the BAO
\begin{eqnarray}
  B_{\rm LPT}^{\rm exact}(k, z)&\simeq&\Biggl[1-{D_1^2(z)\over 1+D_1^2(z)\tilde{g}(k)}
   {\nwP_{22}(k)\over \nwP_{\rm lin}(k)}\Biggr]B_{\rm lin}(k),
\label{lptfit22}
\end{eqnarray}
then we obtain 
\begin{eqnarray}
  B_{\rm LPT}^{\rm fit}(k,z)&=&\Biggl[1-{
   f(\Omega_m h^2,\Omega_b h^2,n_s,\sigma_8,k,z)
\over 1+D_1^2(z)\tilde{g}(k)}
\Biggr]B_{\rm lin}(k),
\label{lptfit}
\end{eqnarray}
as an extended fitting formula. Note that this reduces to the 
expression \eref{leading} in the limit of $\tilde{g}(k)\rightarrow 0$. 
Comparing \eref{lptfit} and \eref{finalfit}, the difference is 
the contribution from $D_1^2(a)\tilde g(k)$ in the denominator 
in front of $\nwP_{22}(k)/\nwP_{\rm lin}(k)$. Since $\tilde g(k)$
is positive, this correction make the damping of the 
BAO signature weaker compared with \eref{finalfit}.
In figure 7, the dashed curve plots $B_{\rm LPT}^{\rm fit}(k,z)$,
\eref{lptfit}.

In order to show the validity of the fitting formula, 
figure 8 plots the relative error s
$|\delta B/B|=|B_{\rm LPT}^{\rm fit}-B_{\rm LPT}^{\rm exact}|
/|B_{\rm LPT}^{\rm exact}|$ at the wavenumber of the peaks and troughs
as a function of the redshift, which is the same as Figure 6 but with 
the LPT instead of the SPT.
As one can see from this figure, the fitting function $B_{\rm LPT}^{\rm fit}$
better reproduces the damping of the BAO signature $B_{\rm LPT}^{\rm exact}$,  
compared with the case of the SPT, especially for larger wavenumber  
even at lower redshift. 
The relative error is less than $10~\%$ in the range of the
wavenumber $k\simlt0.2~{h}{\rm Mpc}^{-1}$ (near the third peak) until
the present epoch, $z=0$. 
In the below, we investigate other possible ways in describing
the damping of the BAO signature due to the nonlinear effect.

One of the other possible formulas for the fitting function is 
the exponential function, as has been discussed in 
references (e.g., \cite{Tegmark,SeoIV,AnguloII,SanchezI}).
Angulo \et ~\cite{AnguloII} discussed the following fitting formula 
with the Gaussian damping function,
\begin{eqnarray}
  B(k,z)=\exp\left[-\left({k\over \sqrt{2}k_*(z)}\right)^2\right]
   B_{\rm lin}(k,z),
\label{angulo}
\end{eqnarray}
where $k_*(z)$ is a time-dependent free parameter,
which should be calibrated by measurements from $N$-body simulations. 
This free parameter describes the time-dependence of the BAO damping, 
and Sanchez \et ~\cite{SanchezI} obtained the best fit value,
$k_*(z)=0.172~h{\rm Mpc}^{-1}$ at $z=1$, from their result 
of $N$-body simulations in \cite{Angulo}.
Their model includes another parameter, $\alpha$, which describes 
the shift of the BAO scale in the wavenumber. However, in
the present paper, we adopt the case of no shift of the BAO scale 
($\alpha=1$) because our interest is focused on the damping of 
the BAO signature  not on the shift of the BAO scale.
It is also shown that $\alpha$ almost equals 1 \cite{Angulo}. 

Figure 9 compares our fitting functions and the Gaussian damping 
function. The vertical axis stands for $B^{\rm fit}(k,z=1)$ divided by 
$B_{\rm lin}(k,z=1)$, which means the damping function. 
The dotted curve and the solid curve are $B_{\rm SPT}^{\rm fit}$
and $B_{\rm LPT}^{\rm fit}$, respectively.
The dashed curve is the Gaussian damping function, \eref{angulo}, with
$k_*(z)=0.172~h{\rm Mpc}^{-1}$.
$B_{\rm SPT}^{\rm fit}$ becomes negative as the wavenumber
becomes larger, $k\simgt0.25~h{\rm Mpc}^{-1}$. This is 
because the phase inversion of the BAO signature  
appears in the SPT, as shown in the panel of $z=1$ in Figure 5, 
where of the SPT breaks down in this regime as mentioned 
in Section 3.2. 
On the other hand, the damping function based on the LPT, 
$B_{\rm LPT}^{\rm fit}/B_{\rm lin}$, shows the similar
behavior as the Gaussian damping function, which approaches 
zero in the limit of $k\rightarrow 0.7$. 
This is due to the modification by the factor, 
$1/[1+D_1^2(z)\tilde g(k)]$, in \eref{lptfit}.
$B_{\rm LPT}^{\rm fit}/B_{\rm lin}$ becomes negative for 
$k\simgt 0.7~h{\rm Mpc}^{-1}$ because of the same reason 
as that for $B_{\rm SPT}^{\rm fit}/B_{\rm lin}$. 

In figure 9, we also plot the following damping function (dot-dashed curve),
\begin{eqnarray}
 B_{\rm SPT}^{\rm exp}(k,z)=\exp\left[-f(\Omega_m h^2,\Omega_b h^2,
n_s,\sigma_8,k,z)\right]B_{\rm lin}(k). 
\label{finalfitb}
\end{eqnarray}
The leading term of the expansion of this damping function 
with respect to $D_1^2(z)$ leads to $B_{\rm SPT}^{\rm fit}$, \eref{finalfit}.
As one can see, this damping function $B_{\rm SPT}^{\rm exp}$ 
agrees with the Gaussian damping function, \eref{angulo}. 
This means that the higher order perturbations
is important in describing the BAO signature for the
regime of the wavenumber, $k\simgt 0.2 h{\rm Mpc^{-1}}$ at redshift $1$.

The result of this section gives us a clue to find how 
to describe the 
BAO damping due to the nonlinear gravitational clustering. 
The BAO damping is determined by the normalization $\sigma_8$
and the growth factor $D_1(z)$.
This suggests that a measurement of the BAO damping might be useful 
to estimate the growth factor of the amplitude of the density 
perturbations, $\sigma_8 D_1(z)$.  
In section 4, we demonstrate a feasibility of constraining 
$\sigma_8 D_1(z)$ by measuring the damping of the BAO signature. 




\section{Feasibility of Constraining  $\sigma_8 D_1(z)$}
As shown in the previous section, the damping of the BAO signature 
is closely related with the amplitude of the power spectrum,
which is determined by $\sigma_8 D_1(z)$, while  the BAO 
signature within the linear theory is determined by 
the density parameters $\Omega_mh^2$ and $\Omega_bh^2$. 
In this section, we discuss the feasibility of constraining 
$\sigma_8 D_1(z)$ by measuring the damping of the BAO in 
the power spectrum in quasi-nonlinear regime. 
To this end, the formula developed in the previous section is useful.

As mentioned in section 1 and subsection 3.3, the redshift-space 
distortions and 
the clustering bias might be additionally influential to the 
damping of the BAO signature. However, we here assume an optimistic 
case that the damping is determined by the quasi-nonlinear clustering 
effect and neglect the effects on the damping of the BAO signature from the 
redshift-space distortions and the clustering bias. 
Then, we study how a measurement of the damping is useful to 
determine $\sigma_8 D_1(z)$. 
Very recently, it is recognized that a measurement of the growth 
factor of the density perturbations is a key to distinguish 
between the dark energy model and modified gravity model for 
the cosmic accelerated expansion (e.g., \cite{YamamotoI,YamamotoII}, 
and references therein).
Our investigation is the first step to investigate if a
measurement of the BAO damping is useful to measure the 
growth factor. 

In our investigation, we adopt a simple Monte Carlo simulation 
of the galaxy power spectrum assuming the $\Lambda$CDM model. 
The error of the galaxy power spectrum depends on its amplitude.
%
For definiteness, we assume the galaxy power spectrum $P_{\rm gal}(k,z)$ 
is modeled with the no-wiggle linear power spectrum in real space, 
$\tilde P_{\rm lin}(k,z)$, as
\begin{eqnarray}
  &&P_{\rm gal}(k, z)=[1+B(k,z)] \tilde P_{\rm gal}(k, z),
\label{redshiftspace}
\end{eqnarray}
with 
\begin{eqnarray}
  &&\tilde P_{\rm gal}(k, z)=b^2{1+Qk^2\over 1+A_1k+A_2k^2}
   \tilde P_{\rm lin}(k, z) ,
\label{redshiftspacenw}
\end{eqnarray}
where we use (\ref{lptfit}) as $B(k,z)$,
$b$ is a constant bias factor, 
and $A_1$, $A_2$ and $Q$ are the parameters, which describe the correction 
of the nonlinear clustering, the redshift-space distortions and the 
scale-dependent bias to the no-wiggle power spectrum.
This model is based on the $Q$-model of Cole \et ~\cite{Cole}, 
and is elaborated by Sanchez \et ~\cite{SanchezI} by adding the new parameter 
$A_2(=Q/10)$ which better reproduces the nonlinear power spectrum at large
wavenumber.
Cole \et ~\cite{Cole} showed that, from numerical study, the value of 
$A_1=1.4$ is adequate to reconstruct the galaxy power spectrum in 
redshift-space, though $Q$ strongly depends on the galaxy
type \cite{Cole,SanchezII,Hamann}. Then, for simplicity, we here 
adopt $A_1=1.4$ and consider the cases $Q=4, ~8, 16, ~32$, to 
estimate the error for the galaxy power spectrum.
We assume the bias parameter $b=2$, for simplicity.

To estimate the constraint on the growth factor, we perform a 
simple Monte-Carlo 
simulation. We assume that the BAO signature, $B(k,z)$, 
can be extracted from a galaxy power spectrum by the method like in the 
reference \cite{PercivalI}.
The variance of the error in measuring the BAO signature
can be estimated by 
\begin{eqnarray}
  \triangle B^2(k)={\triangle P_{\rm gal}^2(k,z)\over [\tilde P_{\rm gal}(k,z)]^2}
\label{variance}
\end{eqnarray}
with
\begin{eqnarray}
  \triangle P_{\rm gal}^2(k,z)=2{(2\pi)^3\over \triangle V_k}{\cal Q}^2(k, z),
\end{eqnarray}
where $\triangle V_k=4\pi k^2\triangle k$ is the volume of the shell in
the Fourier space, and ${\cal Q}^2(k,z)$ is defined as
\begin{eqnarray}
 {\cal Q}^{-2}(k, z)=
\displaystyle{\Delta A} 
\int dz{ds[z]\over dz} s^2[z] {\bar n^2(s[z])
\over [1+\bar n(s[z])P_{\rm gal}(k, s[z])]^2 },
\label{defQ}
\end{eqnarray}
where $\bar{n}(s)$ is the comoving mean number density,
$s=s[z]$ is the comoving distance-redshift relation,
and $\Delta A$ is the survey area.

With the use of the above formulas, we assess $\chi^2$ defined by
\begin{eqnarray}
  \chi^2=\sum_i{\left[B(k_i, z)^{th}-B(k_i, z)^{obs}\right]^2\over 
    \triangle B^2(k_i, z)},
\end{eqnarray}
where $B(k_i, z)^{th}$ is the theoretical one of a fiducial target 
model at the wavenumber $k_i$, while $B(k_i, z)^{obs}$ is the 
corresponding observational one.
The fiducial target model is $\Omega_m=0.28$ and $\sigma_8=0.82$, 
$\Omega_b=0.046$,  $h=0.7$, and $n_s=0.96$, which are the same as those
adopted in figure 1. $B(k_i, z)^{obs}$ is obtained through
a Monte Carlo simulation, following the steps in the below,
\begin{enumerate}
\item 
{Based on the fiducial model, compute $B(k_i, z)$ and the variance 
$\triangle B(k_i, z)$ at $k_i=\triangle k(i-0.5)$, for $i=4,5,\cdots,19$}. 
{Here we specify a bin of the Fourier space,
 $\triangle k=0.01h{\rm Mpc}^{-1}$, and consider the range of 
$0.03<k<0.19$, where the validity of our formula in the previous section is guaranteed.}
We assume two galaxy redshift samples as typical 
future survey. One is the WFMOS-like sample, 
$\Delta A =2000 ~{\rm deg}^2$  in the range of redshift 
$0.5<z<1.3$, and $\bar n=5.0\times 10^{-4}~ [h^{-1}{\rm Mpc}]^{-3}$, 
which contains $2.1\times10^6$ galaxies. 
The other sample assumes the same number density and the range of 
the redshift, but the larger survey area $\Delta A =4\pi$ steradian.

\item 
Each $B(k_i, z)^{obs}$ is obtained through a random process
assuming the Gaussian distribution function with the variance 
$\triangle B^2(k_i, z)$. The data points in figure 10 show
an example of a set of $B(k_i, z)^{obs}$ generated through the random 
process.

\item 
We assess the values of $\chi^2$ with this sample obtained by the
step (i) and (ii).
\end{enumerate}
%
%

We iterate these steps and compute 1000 sets of $\chi^2$, 
then obtain the average of $\chi^2$. 
For the theoretical model, $B(k_i, z)^{th}$, we fixed $\Omega_bh^2$, 
$h$, and $n_s$ as those of the fiducial model, but took 
$\Omega_mh^2$ and $\sigma_8D_1(z=0.9)$ as the variable parameters.
Solid curves in figure 11 show the contour of 
$\triangle \chi^2=2.3$ (inner curve) and $\triangle \chi^2=6.17$ 
(outer curve), which corresponds to the $1~\sigma$ and the 
$2~\sigma$ statistical confidence level, respectively, 
in the $\Omega_mh^2$ and $\sigma_8 D_1(z=0.9)$ plane, 
for the WFMOS-like galaxy sample of $\Delta A =2000 ~{\rm deg}^2$. 
The dotted curves show the same but with the sample of 
the survey area $\Delta A =4\pi$ steradian.
%

{}From figure 11, one can read that 
the 1 $\sigma$ error of $\sigma_8D_1(z=0.9)$ is about 
$0.2$ for the sample of $\Delta A =2000 ~{\rm deg}^2$, 
and is about $0.05$ for the sample of $\Delta A=4\pi$ steradian.
Thus, the constraint on $\sigma_8 D_1(z=0.9)$ of the WFMOS-like 
sample ($\Delta A=2000~{\rm deg}^2$) is not very stringent. 
We already have the small uncertainty of $\sigma_8 D_1(z=0.9)$ 
at the level of a few $\times~0.01$ \cite{Komatsu}, 
from the result by the WMAP observations, which is obtained on 
the basis of the flat CDM 
cosmological model with the cosmological constant. However, we note that
the BAO damping is unique and independent to obtain a constraint on 
$\sigma_8D_1(z)$ around the redshift $1$. 

It would be useful to discuss the origin of the error. 
To this end, we adopt the Fisher matrix approach. As we are considering
the constraint from the damping of the BAO signature, 
we may work with the formula for the Fisher matrix
\begin{eqnarray}
  F_{ij}={1\over 4\pi^2}\int_{k_{\rm min}}^{k_{\rm max}}dk k^2
   {\partial B(k,z)\over \partial \theta_i}
   {\partial B(k,z)\over \partial \theta_j}
   {\nwP_{\rm gal}^2(k,z)\over {\cal Q}^2(k,z)},
\end{eqnarray}
where $\theta_i$ denotes cosmological parameters, for which we focus
on $\sigma_8 D_1(z)$ and $\Omega_mh^2$, $\tilde P_{\rm gal}(k,z)$ is 
defined by (\ref{redshiftspacenw}),
${\cal Q}^2(k,z)$ is defined by (\ref{defQ}), 
we adopt (\ref{lptfit}) as $B(k,z)$, and 
$k_{\min}=0.02h{\rm Mpc}^{-1}$ and $k_{\max}=0.2h{\rm Mpc}^{-1}$.
The minimum error attainable on $\theta_i$
is expressed by the diagonal part of the 
inverse Fisher-matrix, 
$\Delta \theta_i=F_{\theta_i\theta_i}^{-1/2}$,
if the other parameters are known.
Figure 12 plots the errors $\Delta(\sigma_8D_1(z=0.9))$ and
$\Delta(\Omega_mh^2)$, obtained from the Fisher-matrix, as a
function of the mean number density $\bar n(b/2)^2$.\footnote{
Note that the Fisher-matrix depends on the parameter 
$\bar nb^2$.} The thick (thin) solid curve
is  $\Delta(\sigma_8D_1(z=0.9))$ of the sample with 
$\Delta A=2000~{\rm deg.}^2$($\Delta A=4\pi$ radian). 
The dotted curve is $\Delta(\Omega_mh^2)$. 
Here we adopted $Q=16$, but the result is not 
sensitive to this choice. 
One can read that the value of the error
at the point $\bar n(b/2)^2=
5\times 10^{-4}[h^{-1}{\rm Mpc}]^{-3}$ is consistent with the 
result of figure 11. 

Figure 12 demonstrates that the constraint can be improved 
by increasing the number density of the galaxy sample, but
can not be improved for
$\bar n (b/2)^2\simgt 10^{-3}[h^{-1}{\rm Mpc}]^{-3}$. 
The reason can be explained using figure 13, which plots 
$\bar n P_{\rm gal}(k,z=0.9)$ as a function of the wavenumber,
where we fixed $\bar n(b/2)^2=5\times 10^{-4}[h^{-1}{\rm Mpc}]^{-3}$. 
In the region $k<0.2~h{\rm Mpc}^{-1}$,  we have $\bar n P_{\rm gal}(k,z=0.9)
>1$, which means that the shotnoise is not the dominant
component of the error. 
However, the constraint on the parameter is slightly 
improved by increasing the number density $\bar n $,
depending on the bias $b$. 

Finally in this section, we mention the effect 
of the redshift-space distortions and the clustering bias 
on the BAO damping, which might be influential in 
measuring $\sigma_8 D_1(z)$.  It would be true that
there still remains room to investigate how the redshift-space 
distortions and the clustering bias affect the BAO damping. 
If the effect of the redshift-space distortions and the 
clustering bias on the BAO damping could not be clarified, 
it would make it difficult to conclude that the BAO damping
is useful. 
However, very recently, several authors have discussed the 
issue \cite{SanchezI,Matsubara08,KomatsuJeon}. 
If the effect on the BAO damping are well understood, 
it can be a signal, and would be useful in measuring the 
growth factor.

\section{Summary and Conclusions}

In the present paper, we examined the effect of the nonlinear 
gravitational clustering on the BAO signature in the matter 
power spectrum. In particular, we focused on the damping of 
the BAO signature in the quasi-nonlinear regime. 
Our approach is based on the third order perturbation theory 
of the matter fluctuations, which enables us to investigate 
the damping in an analytic way. We found a simple analytic 
expression that describes the damping of the BAO signature, 
which clarifies what the important factor is for the damping. 
We showed that the leading correction for the damping is in 
proportion to the combination of $(\sigma_8 D_1(z))^2$.  
On the basis of the result, we constructed a fitting formula 
for the correction of the damping of the BAO signature in the 
weakly nonlinear regime, which is expressed as a function of 
$k$, $\Omega_mh^2$, $\Omega_bh^2$ and $n_s$. 
This fitting formula reproduces the damping of the BAO 
signature of the second order power spectrum within 
the standard perturbation theory at $10\%$ level for 
$k\simlt0.2~{h}{\rm Mpc}^{-1}$ until $z\sim 1$, though 
the formula is only guaranteed in a narrow range of the 
parameters $\Omega_mh^2$, $\Omega_bh^2$ and $n_s$.

We also discussed a possible extension of our formula to elaborate
higher order nonlinear corrections using a technique 
of resumming infinite series of higher order perturbations 
on the basis of the Lagrangian perturbation theory. 
This extended formula reproduces the damping of the BAO signature of 
the second order power spectrum based on the Lagrangian
perturbation theory at $10\%$ level for 
$k\simlt0.2~{h}{\rm Mpc}^{-1}$ until $z\sim0$. 
This formula was compared with a result of $N$-body simulation, 
which showed the validity of the extended formula.

A measurement of the damping of the BAO signature 
might be useful as a probe of the growth factor of 
the density fluctuations. 
As a first step to investigate such the possibility, we assessed 
the feasibility of constraining $\sigma_8 D_1(z)$ 
by measuring the damping of the BAO in the power spectrum. 
For a useful constraint, we need a very wide survey area of the sky. 
For a definite conclusion, however, we must include the effect of
the redshift-space distortions and the galaxy clustering bias
on the damping of the BAO signature. 
Thus, more sophisticated formula including the effect of the 
redshift-space distortions and the galaxy clustering bias 
at the same time is required. 
If this effect on the BAO damping was well understood, 
it can be useful in measuring the growth factor. 
We plan to revisit this issue in future work.


\ack
We thank A. Taruya, T. Matsubara, Y. Suto and G. H$\ddot{{\rm u}}$tsi 
for useful discussions and comments. 
We are also grateful to anonymous 
referee for useful comments which helped to improve the earlier version of
the manuscript. 
T. N. is supported by a Grand-in-Aid for Japan Society for the Promotion of
Science (JSPS) Fellows (DC1: 19-7066).
This work is supported by Grant-in-Aid
for Scientific research of Japanese Ministry of Education, 
Culture, Sports and Technology (Nos.~18540277, 19035007).

\newpage
\begin{figure}[htbp]
\includegraphics[width=75mm, height=75mm]{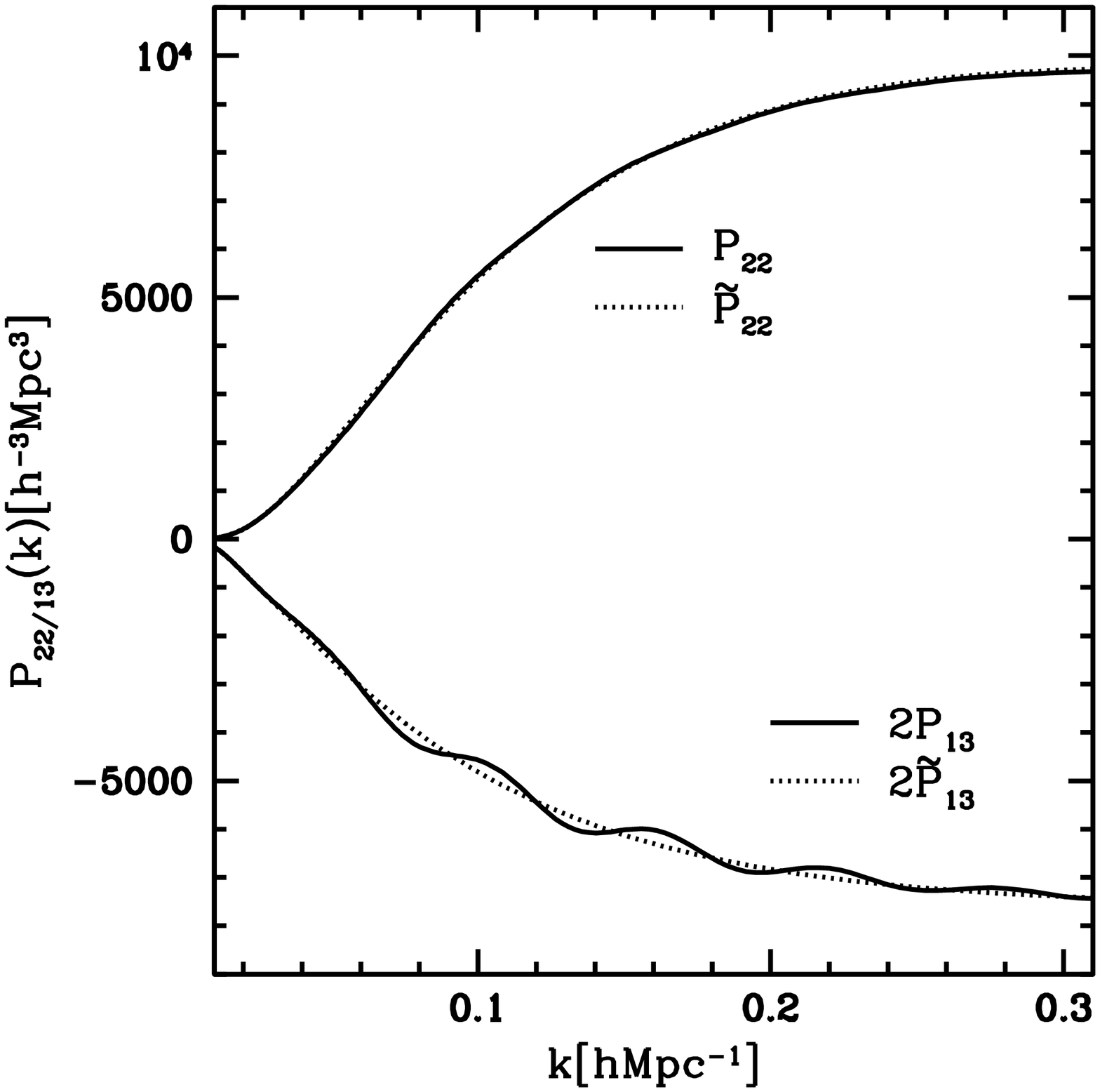}
\caption{Typical behavior of $P_{22}(k)$ (upper solid curve) and
 $2P_{13}(k)$ (lower solid curve), which are calculated using the
 transfer function suggested by \cite{EH} with the BAO. The
 cosmological parameters are $h=0.7$, ${\Omega}_m=0.28$,
 ${\Omega}_b=0.046$, $n_s=0.96$ and $\sigma_8=0.82$. The dotted curves
 show $\tilde P_{22}(k)$ (upper curve) and $2\tilde P_{13}(k)$ (lower
 curve).}
\label{fig1}
\end{figure}

\begin{figure}[htbp]
\includegraphics[width=75mm, height=75mm]{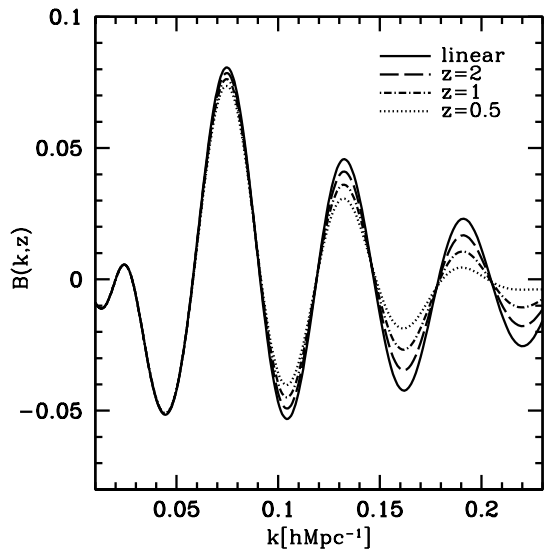}
\caption{$B_{\rm SPT}^{\rm exact}(k,z)$ as a function of $k$ for several
 redshifts, $z=2,~1,~0.5$, which are derived from the matter power
 spectrum including the second order contributions. The solid curve is
 the linear theory. The cosmological parameters are the same as those of
 figure 1. One can see the damping of the amplitude of the BAO
 as the redshift becomes small. In addition, this damping is
 more significant as the wavenumber $k$ is larger.}
\label{fig2}
\end{figure}

\begin{figure}[htbp]
\includegraphics[width=100mm, height=140mm, angle=-90]{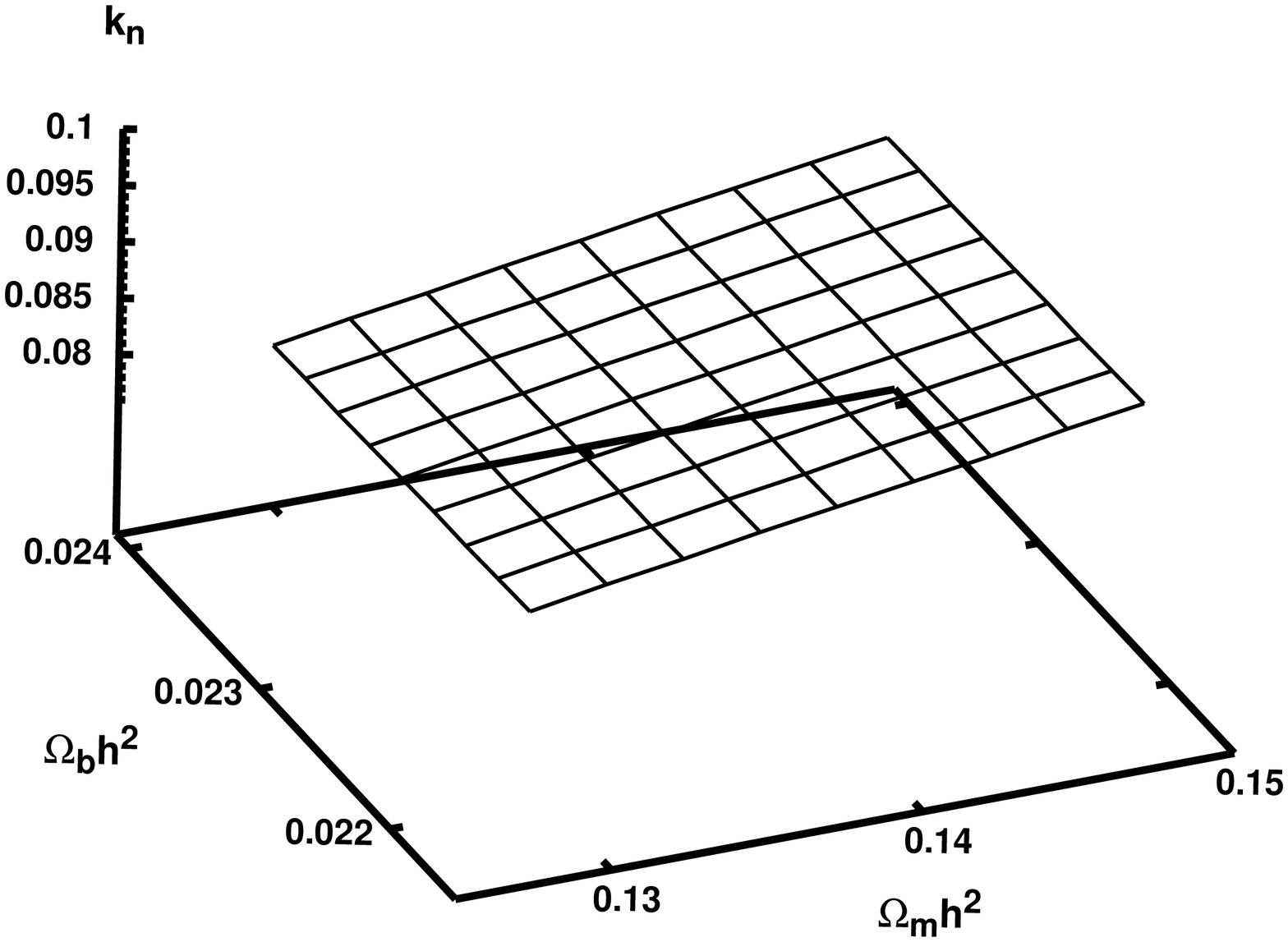}
\caption{The best-fit $k_n$ as a function of $\Omega_mh^2$ and
 $\Omega_bh^2$. The other cosmological parameters are the same as those
 of figure 1.}
\label{fig3}
\end{figure}

\begin{figure}[htbp]
\includegraphics[width=100mm, height=140mm, angle=-90]{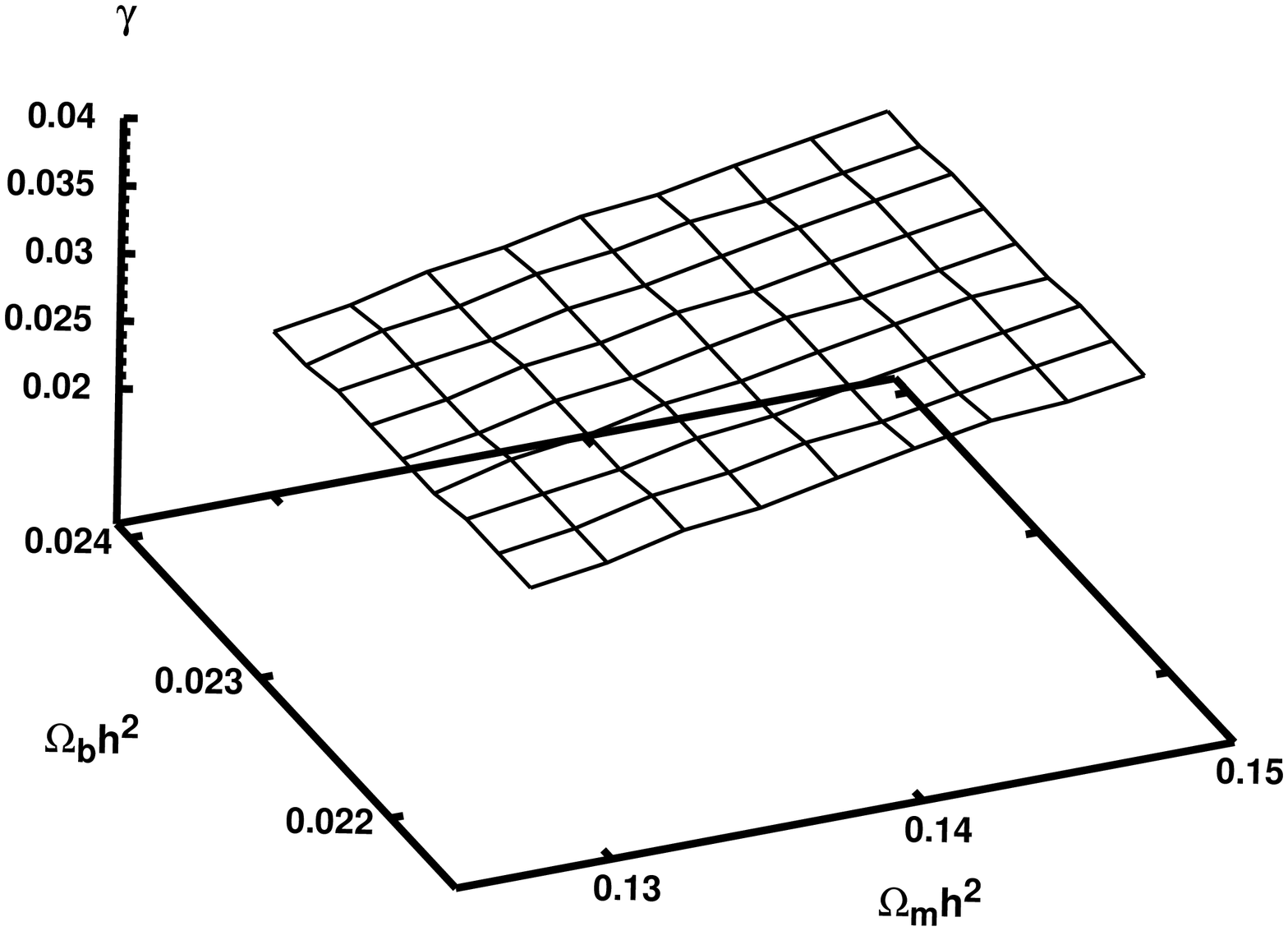}
\caption{The best-fit $\gamma$ as a function of $\Omega_m h^2$ and
 $\Omega_b h^2$.  The other cosmological parameters are the same as
 those of figure 1.}
\label{fig4}
\end{figure}

\begin{center}
\begin{figure}[h]
\includegraphics[width=14cm, height=14cm]{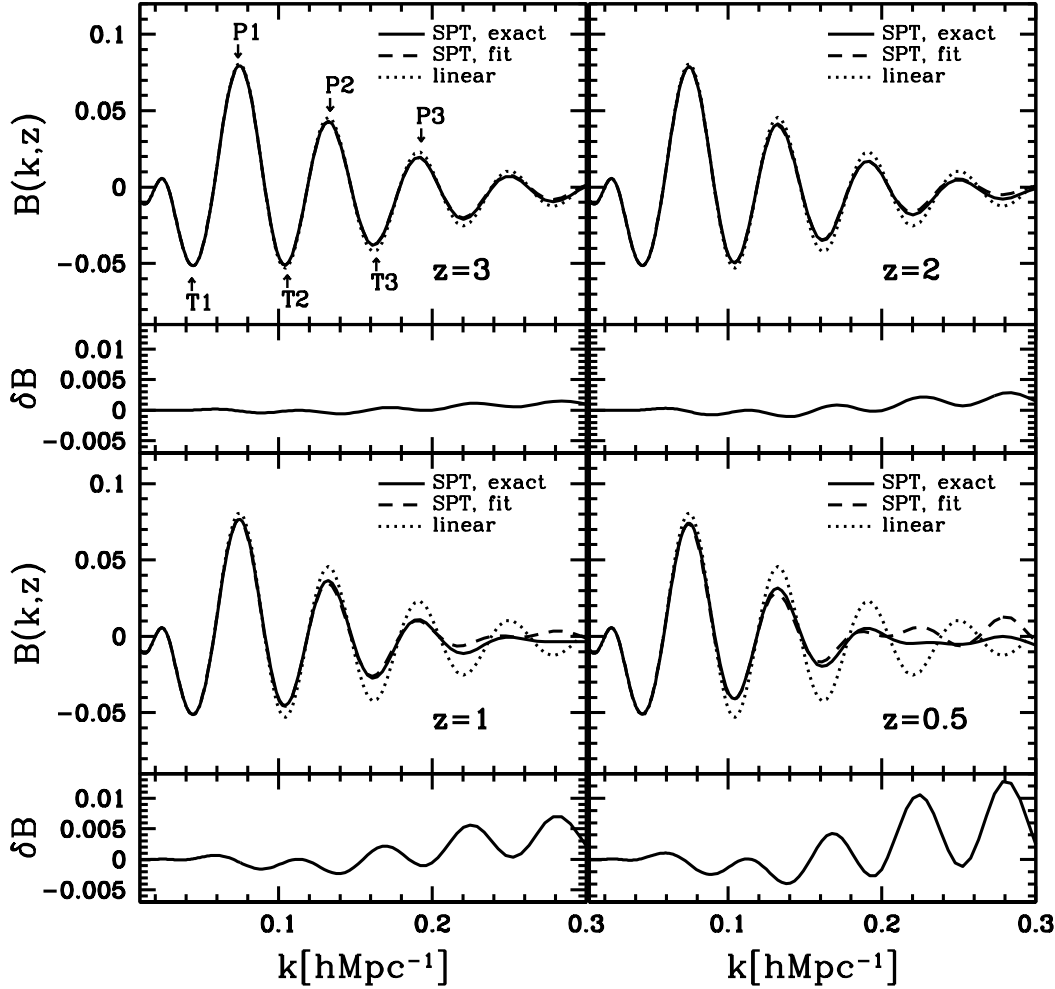}
\caption{ This figure compares our fitting formula
 ($B_{SPT}^{\rm fit}$, dashed curve), (\ref{finalfit}), and the
 formula of the
 third order  perturbation  ($B_{\rm SPT}^{\rm exact}$, solid curve),
 (\ref{defwnl}),  for various redshifts (z=3, 2, 1, 0.5). The dotted
 curve is the linear theory. $\delta B$ is the relative difference
 between the solid curve and the dotted curve 
 ($\delta B=B_{\rm SPT}^{\rm fit}-B_{\rm SPT}^{\rm exact}$). 
 Here, the cosmological parameters are the same as those of Figure
 1. {P1,~P2 and P3 (T1,~T2 and T3) denote the first,  the second and the
 third peak (trough), respectively, used in Figure 6.}
}
\label{fig5}
\end{figure}
\end{center}

\begin{figure}[htbp]
\includegraphics[width=75mm, height=75mm]{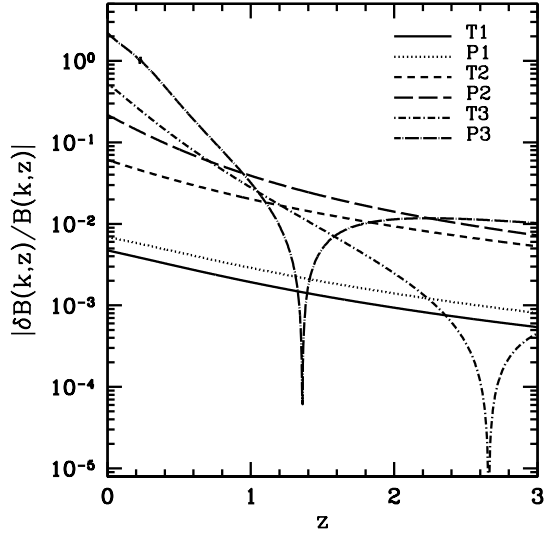}
\caption{{ Relative error
 $|\delta B/B|=|B_{\rm SPT}^{\rm fit}-B_{\rm SPT}^{\rm
 exact}|/|B_{\rm SPT}^{\rm exact}|$ as a function of the redshift at the
 wavenumber of P1,~P2,~P3,~T1,~T2 and T3, respectively, defined in Figure 5. 
}
}
\label{fig6}
\end{figure}

\begin{figure}[htbp]
\includegraphics[width=140mm, height=140mm]{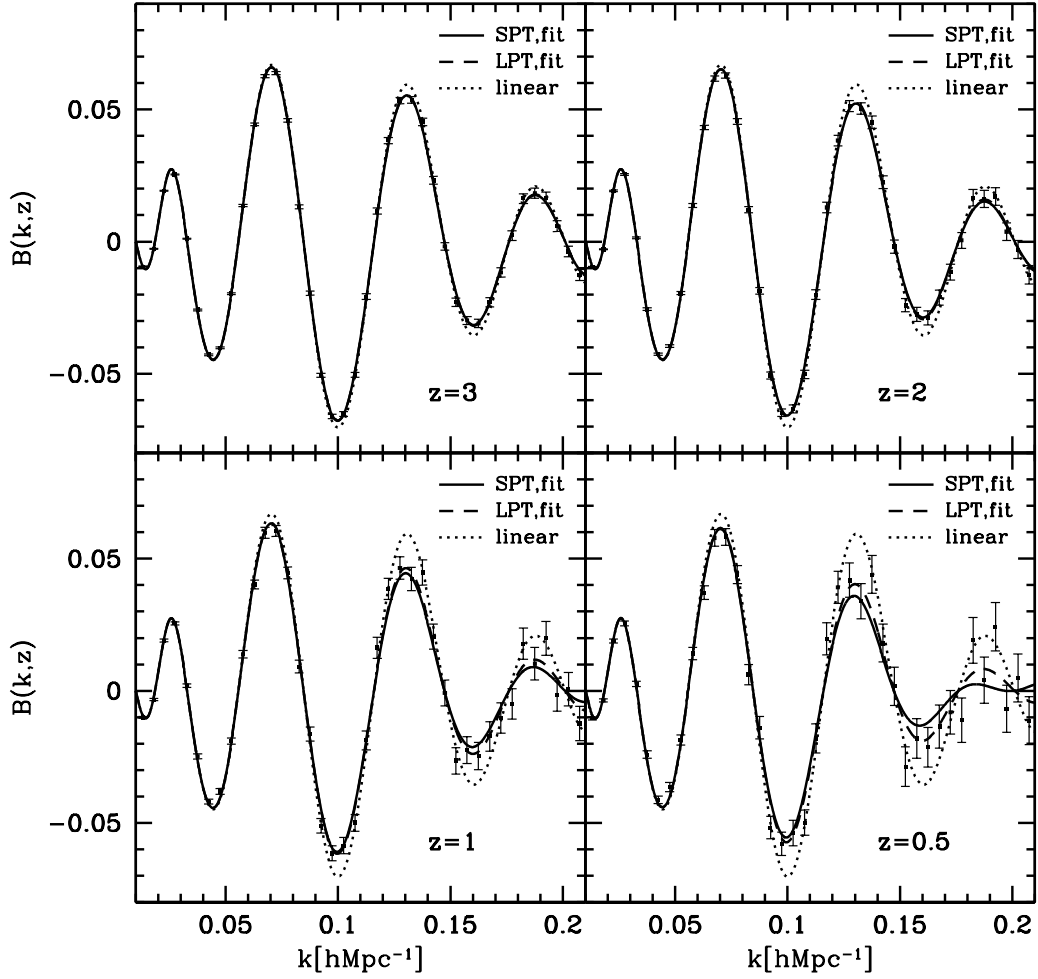}
\caption{The square with the error bar is the result of $N$-body
 simulation. The solid curve is the fitting formula based on the SPT,
 $B_{\rm SPT}^{\rm fit}$ of (\ref{finalfit}), while the dotted
 curve is the linear theory. The dashed curve is the result of an
 extended fitting formula, $B_{\rm LPT}^{\rm fit}$ of
 (\ref{lptfit}). The degree of the damping of $B_{\rm LPT}^{\rm
 fit}$, is slightly weaker than that of $B_{\rm SPT}^{\rm fit}$.
} 
\label{fig7}
\end{figure}

\begin{figure}[htbp]
\includegraphics[width=75mm, height=75mm]{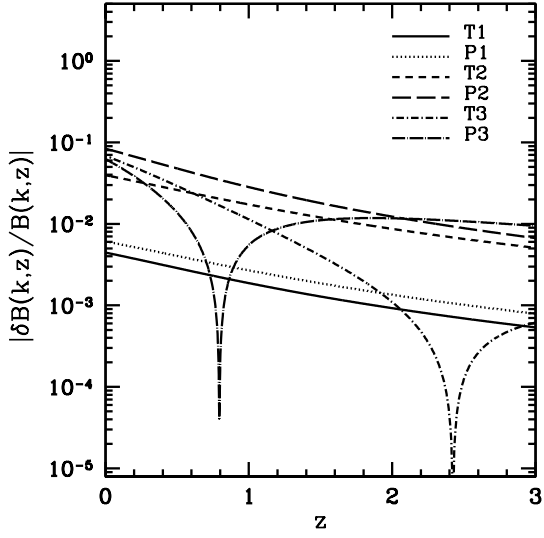}
\caption{Same as Figure 6 but with the LPT instead of the SPT. Relative
 error is
 $|\delta B/B|=|B_{\rm LPT}^{\rm fit}-B_{\rm LPT}^{\rm
 exact}|/|B_{\rm LPT}^{\rm exact}|$.
}
\label{fig8}
\end{figure}

\begin{figure}[htbp]
\includegraphics[width=75mm, height=75mm]{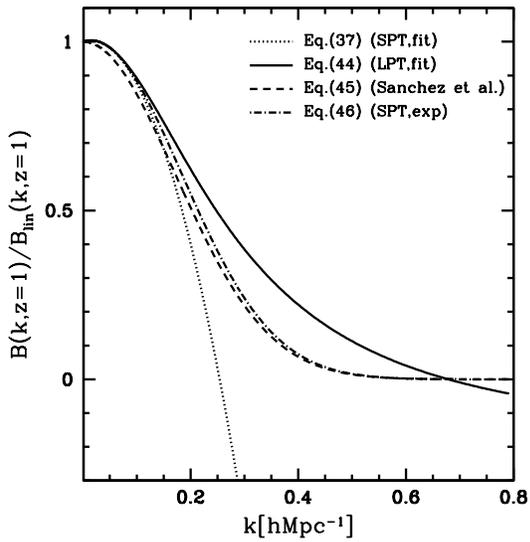}
\caption{Comparison of fitting functions of the damping.  
The dotted curve is $B_{\rm SPT}^{\rm fit}$, \eref{finalfit}, 
while the solid curve is $B_{\rm LPT}^{\rm fit}$, \eref{lptfit}.
The dashed curve is the Gaussian damping function, \eref{angulo}, with
$k_*(z)=0.172~h{\rm Mpc}^{-1}$. The dot-dashed curve is $B_{\rm SPT}^{\rm exp}$, 
\eref{finalfitb}. In this plot, we adopted the same
cosmological parameters as those in \cite{SanchezI}.
}
\label{fig9}
\end{figure}

\begin{figure}[htbp]
\includegraphics[width=75mm, height=75mm]{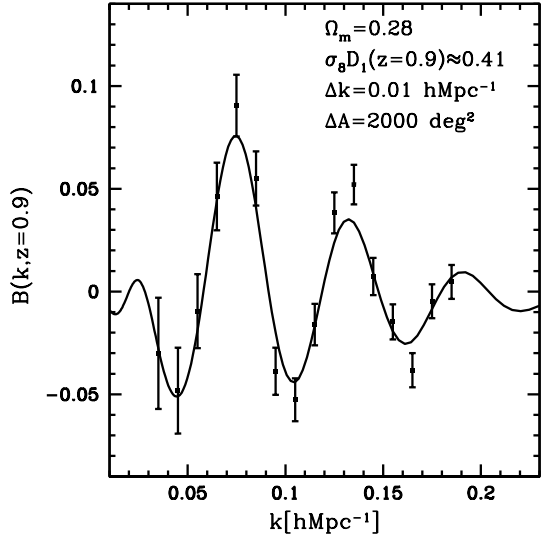}
\caption{
The solid curve is the prediction of the fiducial model, whose
parameters are the same as those of figure 1, while the squares 
with the error bar is an example of $B^{\rm obs}(k_i,z=0.9)$, 
obtained through our Monte Carlo simulation. 
For the galaxy sample, we assumed the WFMOS-like sample 
 of the comoving mean number density $\bar{n}=5.0\times 10^{-4}
 ~[h^{-1}{\rm Mpc}]^{-3}$ and the survey area $2000$ ${\rm deg^2}$ 
 in the range of redshift $0.5<z<1.3$. 
 The bin size of the Fourier space is $\triangle k=0.01~h{\rm Mpc}^{-1}$.} 
\label{fig10}
\end{figure}

\begin{center}
\begin{figure}[htbp]
\includegraphics[width=140mm, height=140mm]{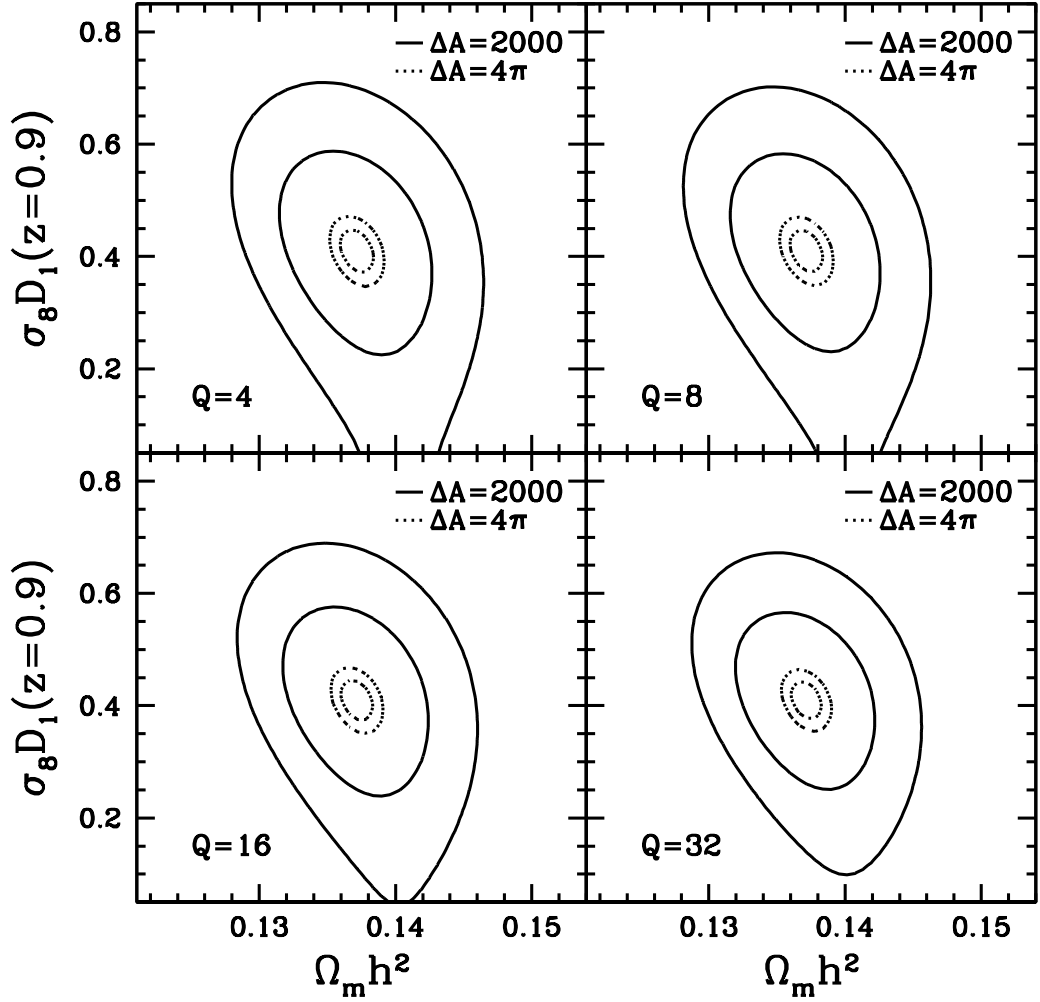}
\caption{Contours of $\triangle \chi^2$ in $\Omega_m h^2$ and
 $\sigma_8 D_1(z=0.9)$ plane. 
 The solid curves assume the WFMOS-like sample of 
 the survey area,  $\Delta A=2000$ ${\rm deg^2}$. 
 Inner (outer) curve is the contour of $\triangle \chi^2=2.3$
 ($\triangle\chi^2=6.17$), which corresponds to $1~\sigma$ ($2~\sigma$)
 confidence level.
 The dotted curve is the same as the solid curve, but assumes the 
survey area, 
 $\Delta A=4\pi$ steradian.
}
\label{fig11}
\end{figure}
\end{center}

\begin{figure}[htbp]
\includegraphics[width=75mm, height=75mm]{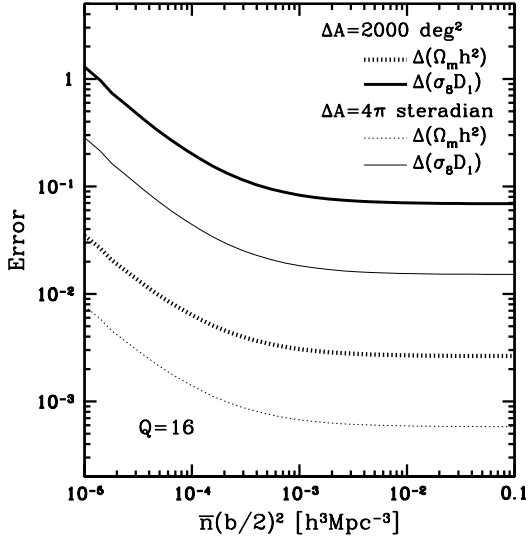}
\caption{The $1\sigma$-level statistical errors of $\Omega_mh^2$ (solid
 curve) and $\sigma_8D_1(z=0.9)$ (dotted line) as a function of the number
 density $\bar n (b/2)^2$. 
The thick curves assume $\Delta A=2000~{\rm deg}^2$, and the thin
 curves assume $\Delta=4\pi$ steradian. In this figure, we fixed $Q=16$.
} 
\label{fig12}
\end{figure}

\begin{figure}[htbp]
\includegraphics[width=75mm, height=75mm]{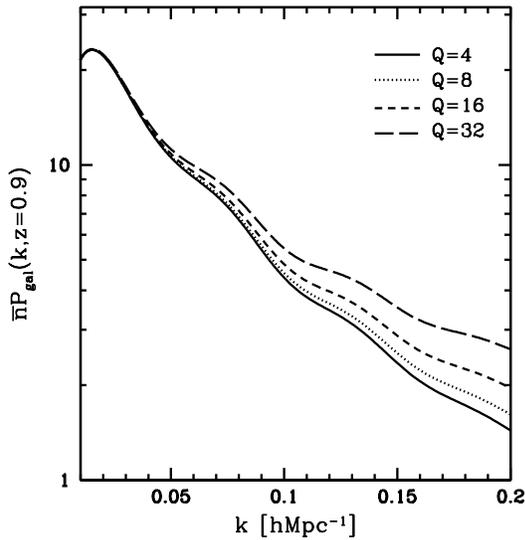}
\caption{The galaxy power spectrum multiplied by the mean number 
density, $\bar nP_{\rm gal}(k,z=0.9)$,  as a
 function of $k$. Here, we fixed $\bar n (b/2)^2=5\times 10^{-4}
~[h^{-1}{\rm Mpc}]^{-3}$. We have $\bar nP_{\rm gal}(k,z=0.9)>1$ for 
 $k< 0.2~h{\rm Mpc}^{-1}$, where the shotnoise is subdominant.} 
\label{fig13}
\end{figure}

\end{document}